\documentclass[twocolumn]{aastex7}
\usepackage[T1]{fontenc}
\usepackage{nameref}

\usepackage{CJKutf8}

\newif\ifshowedits
\showeditstrue   

\begin{document}

\title{Searching for Circumplanetary Dust in the Cavity of the AB Aurigae Disk with ALMA}

\author[0000-0003-4006-102X]{Lillian Yushu Jiang \begin{CJK*}{UTF8}{gbsn}(蒋雨舒)\end{CJK*}}
\email{lyjiang@ucsb.edu}
\affiliation{Department of Physics, University of California, Santa Barbara, Santa Barbara, CA 93106, USA}

\author[0000-0003-2649-2288]{Brendan P. Bowler}
\affiliation{Department of Physics, University of California, Santa Barbara, Santa Barbara, CA 93106, USA}
\email{bpbowler@ucsb.edu}

\author[0000-0002-4392-1446]{Ya-lin Wu}
\affiliation{Department of Physics, National Taiwan Normal University, Taipei 116, Taiwan}
\email{yalinwu@ntnu.edu.tw}

\author[0000-0003-2969-6040]{Yifan Zhou}
\affiliation{Department of Astronomy, University of Virginia, Charlottesville, VA 22904, USA}
\email{yzhou@virginia.edu}

\received{Sep 19, 2026}
\accepted{Oct 26, 2026}

\begin{abstract}
The young intermediate-mass star AB Aurigae is surrounded by a transition disk whose cavity contains several proposed protoplanets, including the debated source ``AB~Aur~b" at a projected separation of ${\approx}0\farcs6$ (${\approx}93$\,au). To search for circumplanetary material associated with these candidates, we have obtained the most sensitive (${\sim}20\,\mu$Jy\,beam$^{-1}$ rms) and highest angular resolution (${\sim}30$\,mas) ALMA Band~7 (0.88\,mm) observations of AB~Aur to date. We recover the asymmetric dust ring in the continuum emission peaking at ${\sim}0\farcs9$ (${\sim}140$\,au) and compact emission at the stellar position, but detect no point source at the location of AB~Aur~b or any of the other proposed protoplanet candidates down to a $3\sigma$ limit of $60\,\mu$Jy. For optically thin emission, this corresponds to a dust mass of $\lesssim 0.0059\,M_\oplus$ (for a dust temperature of 40~K), comparable to the circumplanetary disk around PDS~70~c. The $^{12}$CO emission fills the continuum cavity, revealing inner spiral structure and a clear rotation pattern. The gas at the position of AB~Aur~b is smoothly connected to the ambient disk with no localized excess or obvious kinematic deviation. If any proposed candidates are accreting, their circumplanetary material must have a low dust mass, a small physical size, dust that is optically thick, or a dearth of large mm-emitting grains.
\end{abstract}

\section{Introduction} \label{sec:intro}
Directly imaging forming planets embedded in protoplanetary disks  offers one of the most direct ways to study planet assembly, yet  robust detections at optical- and near-infrared wavelengths remain rare as a result of extreme star--planet contrasts, speckle  noise, and confusion with disk substructure \citep{Bowler2016, Follette2023}. Indeed, several early candidate detections illustrate how disk structure can mimic protoplanet emission. For example, \citet{Currie2019} and \citet{Sallum2023} argued that the LkCa~15 signals reported by \citet{Kraus2012} and \citet{Sallum2015} were consistent with inner-disk emission; \citet{Cheetham2015} attributed the T~Cha companion signal reported by \citet{Huelamo2011} to forward scattering from the disk rim; and \citet{Follette2017} found that the candidate HD~100546~c \citep{Brittain2013, Currie2015} was an artifact of aggressive processing of a disk feature.

Beyond detecting emission from the planet, characterizing the circumplanetary material that surrounds and feeds nascent giant planets presents a further observational challenge. Circumplanetary disks (CPDs) trace ongoing mass and angular-momentum transfer onto young giant planets, helping set their final masses \citep{Lubow1999,Tanigawa2012,Szulagyi2014}. Their thermal structure governs their rotational support \citep{Krapp2024} and observable emission \citep{Choksi2025,Taylor2025,Taylor2026}. CPD properties and dispersal timescales also shape the conditions for satellite formation and survival \citep{Canup2002}. Moving to millimeter wavelengths can provide a clearer view of circumplanetary material by tracing cool dust and molecular gas while mitigating extinction and the strong scattered-light signatures that dominate at shorter wavelengths \citep[e.g.,][]{Wu2020}.

ALMA has been the workhorse facility for this effort. Millimeter continuum emission has been detected around three planetary-mass objects in different environments: the isolated object OTS~44 \citep{Bayo2017}, the wide-orbit companion SR~12~c \citep{Wu2022}, and the embedded protoplanet PDS~70~c \citep{Isella2019,Benisty2021}. Non-detections are equally valuable; \citet{Bowler2015} limit the circumplanetary dust mass of the accreting companion GSC~6214-210~B to $<0.15\ M_\oplus$. \citet{Andrews2021} systematically searched gaps in the DSHARP disks, reaching dust-mass sensitivities of $\sim0.001$--$0.2\ M_\oplus$. Deep observations by \citet{Facchini2026} and \citet{Ricci2017} also yielded stringent limits on circumplanetary emission from WISPIT~2b and 2M1207~b, respectively, illustrating how CPDs may remain undetected if their millimeter-emitting material is depleted or highly compact.

\section{AB Aurigae and its Protoplanet Candidates} \label{sec:ab_aur}
AB~Aurigae (AB~Aur) is a young ($\approx$2--4~Myr), accreting A0 Herbig~Ae star in the Taurus--Auriga star-forming complex \citep{dewarf2003}, at a Gaia DR3 distance of $155.9\pm0.9$~pc \citep{Bailer-Jones2021}. Its circumstellar environment has been studied for decades from optical to millimeter wavelengths \citep[e.g.,][]{Sanford1958,Mannings1997,Grady1999,Fukagawa2004,Pietu2005,Speedie2024}, revealing a large, nearly face-on, highly structured protoplanetary disk with pronounced non-axisymmetric substructure. The system hosts a small flared inner disk ($\sim$2--5~au) \citep{difolco2009}, surrounded by a broad central cavity with a radius of $\sim$100~au and a bright dust ring seen in millimeter continuum emission \citep{Pietu2005}. Within this cavity, \citet{tang2017} resolved two prominent CO spiral structures. On larger scales, \citet{Speedie2024} traced outer spiral arms extending to at least $\sim$450~au and interpreted kinematic signatures in ALMA isovelocity maps as evidence for gravitational instability. Follow-up $^{12}$CO and SO observations by \citet{Speedie2025} identified three infalling streamers, indicating that late-stage accretion is replenishing the disk and may help sustain its gravitational instability.

Several candidate young planets have been reported in the AB~Aur system. \citet{Dong2016} proposed a $3\,M_{\rm Jup}$ gap-opening planet at 67~au ($\sim$0\farcs4) to reproduce the gap observed in scattered light. Using ALMA CO observations, \citet{tang2017} subsequently inferred two perturbers at 0\farcs21 (T17~P1) and 0\farcs57 (T17~P2) from the inner spiral morphology and dust-ring edge, estimating a mass of $\approx14\,M_{\rm Jup}$ for T17~P1. VLT/SPHERE near-IR polarimetry revealed two compact features at 0\farcs18 (B20~f1) and 0\farcs67 (B20~f2) \citep{Boccaletti2020}. Using Subaru/SCExAO-CHARIS and HST/STIS coronagraphic imaging, \citet{Currie2022} reported AB~Aur~b at 0\farcs57, and two more sources at 2\farcs8 and 3\farcs7. Using IRTF/iSHELL spectroscopy and spectroastrometry, \citet{Kozdon2026} detected an additional compact $^{12}$CO-emitting source at 0\farcs41 (K26~CO). Finally, \citet{Boccaletti2026} noted a third near-IR polarized-light feature at 0\farcs18 (B26~f3) and a tentative H$\alpha$ signal at 0\farcs35 (B26~H$\alpha$).

Among these, AB~Aur~b is particularly intriguing but has attracted considerable scrutiny. \citet{Currie2022} interpreted the source as a protoplanet based on its resolved morphology and multiwavelength emission: it is spatially extended relative to a point source, shows orbital motion over a 13-yr baseline, has a lower near-infrared polarization fraction (${\lesssim}20\%$) than the surrounding disk, and has red near-infrared colors ($J-K_s=1.1\pm0.3$ mag). Overall, its SED can be reproduced by a ${\sim}2000$--$2500$\,K atmosphere, which would imply a mass in the planetary regime. They noted that it lies interior to the millimeter dust ring, at the location predicted for a perturber driving the inner CO spirals \citep{tang2017}, and that most orbital solutions do not cross the ring.

Using HST/WFC3 H$\alpha$ imaging with the F656N filter, \citet{Zhou2022} recovered emission at the location of AB~Aur~b but measured a line-to-continuum ratio similar to that of the host star. In subsequent HST/WFC3 UV-to-optical imaging with the F336W, F410M, and F645N filters, \citet{zhou2023} found the source’s colors to be consistent with scattered stellar light rather than planetary photospheric or accretion-shock emission. \citet{Biddle2024} did not detect Pa$\beta$ emission with Keck/NIRC2, suggesting that, if the source is planetary, its accretion was weak or stochastic during those observations. Using five epochs of HST/WFC3 H$\alpha$ imaging, \citet{Bowler2025} found that AB~Aur~b varied much more strongly than the star and showed no correlated variability. This ruled out unobstructed scattered starlight as the sole origin, but did not rule out more complex disk-related explanations, including shadowing from inner-disk features. \citet{Currie2025} used VLT/MUSE spectroscopy to detect a H$\alpha$ profile resembling an inverse P~Cygni profile, which they interpreted as evidence for infalling gas. \citet{Boccaletti2026} did not detect AB~Aur~b in multi-epoch VLT/SPHERE-IRDIS polarimetry and recovered only a marginal signal at its position in SPHERE/ZIMPOL H$\alpha$ imaging. Most recently, \citet{Lewis2026} obtained the first $L^\prime$-band polarized-light images of the disk with Keck/NIRC2-Pol; they resolved the inner spiral arms but found no enhanced polarized emission at the location of AB~Aur~b. Overall, the available observations do not yet uniquely distinguish a planetary source from emission associated with the structured circumstellar disk.

Here, we present ALMA Band~7 (0.88\,mm) observations of the AB~Aur system to search for compact dust emission from a potential CPD around AB~Aur~b and the other proposed candidates. This study is organized as follows. Section~\ref{sec:obs} describes the observations and data reduction. Section~\ref{sec:result} presents the 0.88\,mm continuum images and the $^{12}$CO $J=3$--2 emission, and reports our limits at the candidate locations. Section~\ref{sec:discussion} places these limits in the context of disks around stars and other planetary-mass companions, and Section~\ref{sec:summary} summarizes our conclusions.

\begin{figure*}[!t]
    \centering
    \includegraphics[width=\linewidth]{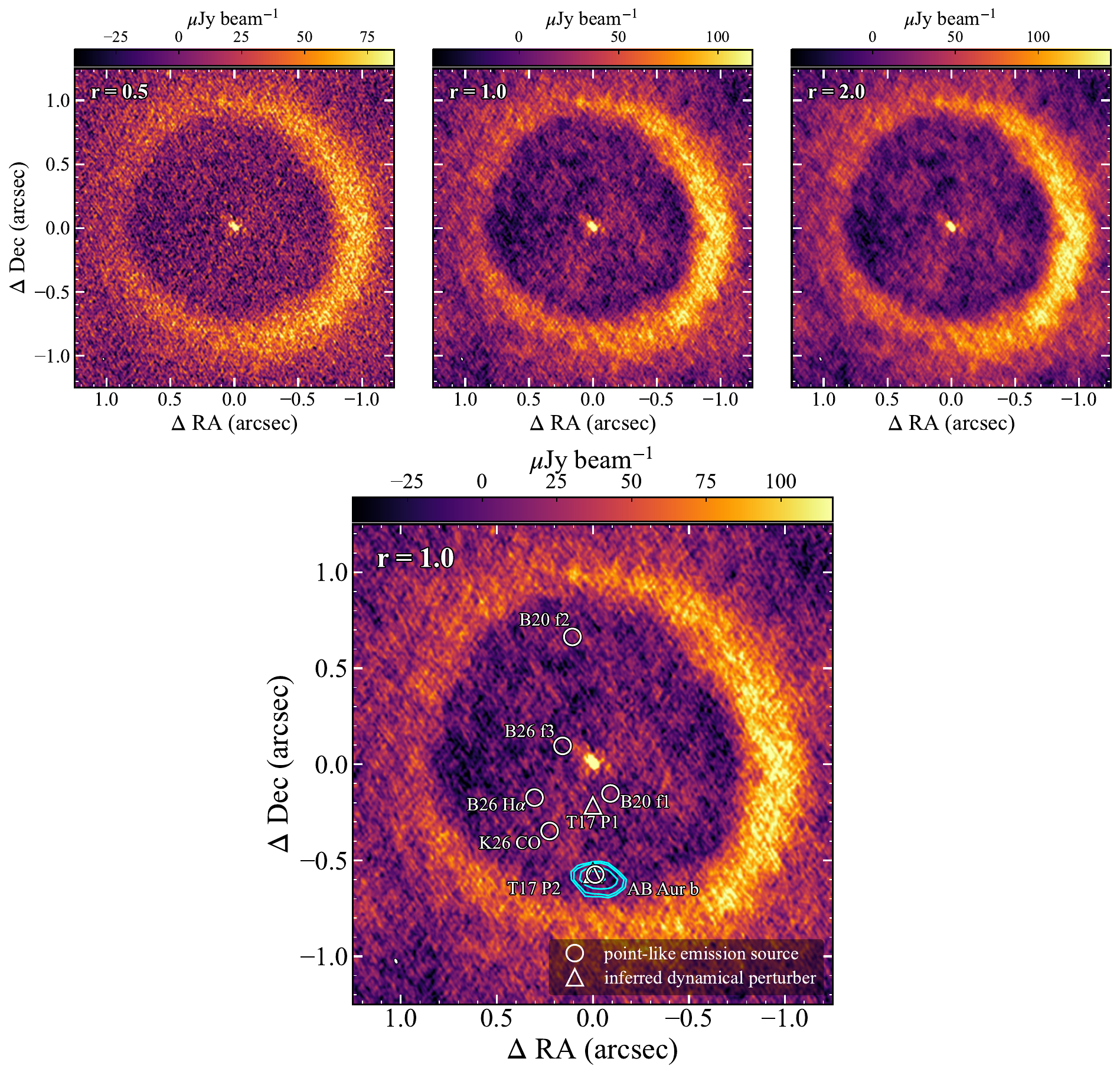}
    \caption{ALMA 0.88~mm continuum images of the AB~Aur disk produced with Briggs robust values ($r$) of 0.5, 1.0, and 2.0 (top row). The bottom panel shows the $r$=1.0 image, with cyan contours tracing the HST/WFC3 H$\alpha$ detection of AB~Aur~b at 8, 10, 15, 20, and $25\times$ the image rms from \citet{Bowler2025}. Open symbols mark planet candidates within the field (Section~\ref{sec:ab_aur}): circles denote point-like emission sources and triangles denote dynamically inferred perturbers, following \citet{Speedie2024}. The dust ring peaks at $\sim0\farcs9$ ($\sim140$\,au), and compact emission is detected at the stellar position; however, no significant point-source emission is detected at AB~Aur~b or the other candidate locations. The synthesized beam is shown in the lower-left corner of each panel.}
    \label{fig:ab_aur_mm_halpha}
\end{figure*}

\section{Observations and Data Reduction} \label{sec:obs}
The ALMA Band~7 observations of AB~Aur were obtained under project 2022.1.01360.S (PI: Bowler) on UT 2023 August 16, August 21, and September 11, comprising three of eight scheduled execution blocks. (The September block was aborted early but retained after passing QA0). The 12~m array configuration used 42--51 antennas with baselines from 67~m to 14.9~km, giving a maximum recoverable scale of ${\sim}1\farcs3$--$1\farcs6$ at 340~GHz. The total on-source integration time on AB~Aur is 106~min. Per-execution details are listed in Table~\ref{tab:alma_obs}.

The correlator setup comprises three 2.0~GHz continuum windows centered near 333.8, 335.8, and 347.8~GHz, and a fourth higher spectral resolution window covering the $^{12}$CO $J=3$--2 transition at 345.796~GHz. The data are calibrated with the standard ALMA pipeline (CASA v6.5.4.9, pipeline v2023.1.0.124; \citealt{casa2022}), using the quasars J0510+1800 as the bandpass and flux calibrator, J0438+3004 as the phase calibrator, and J0440+2728 as a check source.

For the continuum, we exclude the $^{12}$CO line channels, yielding an effective bandwidth of ${\approx}5.9$~GHz at a representative frequency of 340.8~GHz. We image the combined visibilities with \texttt{tclean} in multifrequency-synthesis mode, applying multiscale \textsc{clean} and one round of phase-only self-calibration on the continuum model \citep{tclean}. We produce images at Briggs robust $0.5$, $1.0$, and $2.0$, and adopt the robust $1.0$ image as our fiducial result, with a synthesized beam of $32.1\times18.9$~mas (${\rm PA}=23\fdg4$) and an off-source rms of ${\approx}20\,\mu$Jy\,beam$^{-1}$ measured outside the disk emission.

For the $^{12}$CO $J=3$--2 line, we image the continuum-subtracted spectral window with the same parameters as the continuum at robust of $1.0$, producing a cube with 40 channels of width 0.488~MHz ($\approx0.42$~km\,s$^{-1}$) and a synthesized beam of $31.8\times18.6$~mas (${\rm PA}=23\fdg5$).





\begin{deluxetable}{ccccccc}
\tablecaption{ALMA Band 7 Observations of AB Aur\label{tab:alma_obs}}
\tablehead{
\colhead{EB Date} & 
\colhead{Start--End (UTC)} & 
\colhead{$N_\mathrm{ant}$} & 
\colhead{Baseline (m)} & 
\colhead{$T_\mathrm{on}$ (min)} & 
\colhead{PWV (mm)} &
\colhead{$\phi_\mathrm{rms}$ (deg)}
}
\startdata
2023 Aug 16 & 09:56--11:57 & 46 & 83--14851  & 45.7 & 0.5 & 15.0 \\
2023 Aug 21 & 08:45--10:46 & 42 & 83--14851  & 44.8 & 0.5 & 13.8 \\
2023 Sep 11 & 10:28--11:19 & 44 & 67--12272  & 15.4 & 0.5 & 15.9 \\
\enddata
\tablecomments{The 2023 September 11 execution block was aborted  early but retained after passing QA0 inspection. The total  on-source integration time across all three blocks is 105.9~min.  $N_\mathrm{ant}$ is the number of antennas after flagging as  reported in the QA2 report. $\phi_\mathrm{rms}$ is the  WVR-corrected phase rms averaged over all basebands and scans.  J0510+1800 served as the bandpass and flux calibrator, J0438+3004  as the phase calibrator, and J0440+2728 as a check source for  all three executions.}
\end{deluxetable}

\section{Results} \label{sec:result}

\subsection{0.88 mm Continuum} \label{sec:continuum}

Figure~\ref{fig:ab_aur_mm_halpha} shows the 0.88\,mm continuum images at Briggs robust values of 0.5, 1.0, and 2.0. We recover the well-known dust ring, which peaks at a radius of ${\sim}0\farcs9$ (${\sim}140$\,au) from the host star and shows a pronounced azimuthal brightness asymmetry, with the western to southwestern side of the ring substantially brighter than the eastern side \citep[e.g.,][]{Pietu2005,tang2017}. With Briggs weighting ($\mathrm{robust}=1.0$), our image achieves a synthesized beam of $32.1\times18.9$~mas, providing the highest angular resolution yet reported for AB~Aur at 0.88~mm.

We also detect compact continuum emission at the stellar position with a peak intensity of $1.2$\,mJy\,beam$^{-1}$ (${\sim}60\sigma$) and an integrated flux density of $3.12\pm0.45$~mJy within
a $0.1''$-radius aperture, with the uncertainty including
local aperture scatter and a 10\% flux-calibration uncertainty \citep{privon_2022_7662388}.
Our image marginally resolves the central source, with a beam-deconvolved size of approximately $23\times11$~mas ($3.5\times1.7$~au) at a position angle of ${\sim}30^\circ$. \citet{RM2024} and \citet{Rota2024} analyzed archival ALMA Band~7 observations (project 2012.1.00303.S; PI Y.-W. Tang) but recovered different fluxes for the central source: the former measured $3.3\pm0.5$~mJy by integrating within the $5\sigma$ contour, while the latter obtained $1.31\pm0.29$~mJy from a two-dimensional Gaussian fit. This discrepancy likely reflects the different flux-extraction methods. \citet{RM2024} estimated that free--free emission contributed roughly one-third of the central flux, with the rest attributed to dust. The central emission therefore likely contains both free--free and thermal dust components \citep{difolco2009, Francis2020, Rodriguez2014}.

We detect no significant point source at the location of AB~Aur~b, at any of the other proposed candidates, or elsewhere interior to the ring (Figure~\ref{fig:ab_aur_mm_halpha}). This includes T17~P1 and B20~f1, which lie near compact $^{12}$CO emission (Section~\ref{sec:co}). The local rms noise of $\sigma = 20\,\mu$Jy\,beam$^{-1}$ sets a $3\sigma$ upper limit of $60\,\mu$Jy on point-source emission. For comparison, the archival Band~7 data from project 2012.1.00303.S presented in \citet{RM2024} reached an rms of $200\,\mu$Jy\,beam$^{-1}$ with a $0.362''\times0.228''$ beam. Our observations are ten times more sensitive to point sources at a much finer resolution of $0.032''\times0.019''$.

The high-resolution 0.88\,mm image also reveals a potential radial widening at the southern end of the bright western arc, near $(\Delta{\rm RA},\Delta{\rm Dec})\approx(-0\farcs5,-0\farcs7)$, where the ring’s inner edge protrudes toward the cavity. Previous observations revealed the large-scale asymmetry \citep[e.g.,][]{Tang2012} but did not resolve this feature. We do not recover the possible weak feature extending from the central source to the southwestern rim of the disk identified by \citet{RM2024}.

Assuming the continuum emission is optically thin, the dust mass of a putative CPD around any young planets within the AB~Aur transition disk can be derived following
\citet{Hildebrand1983}:
\begin{equation}
M_{\rm dust} = \frac{F_\nu \, d^2}{\kappa_\nu \, B_\nu(T_{\rm dust})},
\end{equation}
where $d$ is the distance to AB~Aur, $\kappa_\nu$ is the dust opacity at the observing frequency $\nu$, and $B_\nu(T_{\rm dust})$ is the Planck function at the assumed dust temperature.

We adopt a distance of $d = 155.9$\,pc \citep{gaiadr3} and a dust opacity of $\kappa_\nu = 3.4\,{\rm cm^2\,g^{-1}}$ at 880\,$\mu$m, computed from the frequency-dependent opacity relation $\kappa_\nu = 10\,(\nu / 10^{12}\,{\rm Hz})\,{\rm cm^2\,g^{-1}}$ of \citet{Beckwith1990}.

Following \citet{Isella2014} and \citet{Keppler2019}, the CPD dust temperature is described by
\begin{equation}
T_{\rm dust}^4(r)=T_{s,{\rm irr}}^4(a_p)+T_{p,{\rm irr}}^4(r)+T_{\rm acc}^4(r),
\end{equation}
where the terms represent stellar irradiation ($T_{s,{\rm irr}}$), planetary irradiation ($T_{p,{\rm irr}}$), and viscous dissipation through accretion ($T_{\rm acc}$), respectively; $a_p$ and $r$ are the star--planet and planet--dust distances. Stellar irradiation dominates the heating budget that \citet{Benisty2021} estimated for the CPD around PDS~70~c: $T_{s,{\rm irr}}=24$~K, $T_{p,{\rm irr}}=18$~K, and $T_{\rm acc}=3$~K, giving $T_{\rm dust}\approx26$~K. Neglecting local circumplanetary heating, we conservatively adopt $T_{\rm dust}\sim\,T_{s,{\rm irr}}=40$~K, consistent with the midplane temperature near 100~au in the passive AB~Aur disk model of \citet{Tannirkulam2008}. This also agrees with the approximately 42~K midplane temperature adopted near this radius by \citet{RM2020}, whose profile was calibrated to molecular-line temperatures farther out in the disk. For optically thin emission, the $3\sigma$ flux limit of $60\,\mu$Jy implies
\begin{equation}
M_{\rm dust}\lesssim0.0059\,M_\oplus\approx0.48\,M_{\rm Moon}.
\end{equation}
Additional heating from planetary irradiation and accretion shocks would lower this mass limit. Additional heating from planetary irradiation and viscous accretion would lower this mass limit. For the best-fit parameters of \citet{Currie2022} ($9\,M_{\rm Jup}$, $2.75\,R_{\rm Jup}$, $T_{\rm eff}=2200$~K, $\dot{M}\approx10^{-6}\,M_{\rm Jup}\,{\rm yr^{-1}}$), we scale $T_{p,{\rm irr}}$ from the PDS~70~c value of 18~K by $(L_p/L_{\rm PDS\,70\,c})^{1/4}$, where the photospheric luminosity $L_p=4\pi R_p^2\sigma T_{\rm eff}^4$ of AB~Aur~b is ${\approx}36$ times that of PDS~70~c, and compute $T^4_{\rm acc}=3GM_p\dot{M}/(8\pi\sigma r^3)$ directly. This gives $T_{p,{\rm irr}}\approx44$~K and $T_{\rm acc}\approx15$~K at 1~au from the planet, falling to ${\approx}24$ and ${\approx}6$~K at the CPD truncation radius ($R_{\rm H}/3\approx3.3$~au). The resulting $T_{\rm dust}\approx41$--50~K lowers the dust-mass limit by only 3--24\%, to $0.0045$--$0.0057\,M_\oplus$.

For a colder dust temperature of $T_{\rm dust} = 15$\,K, as is commonly adopted for faint disks around isolated planetary-mass objects \citep[e.g.,][]{Wu2020}, the corresponding limit is $M_{\rm dust} \lesssim 0.023\,M_\oplus$. We place these limits in the context of disks around stars and planetary-mass companions in Section~\ref{sec:discussion}.

\subsection{$^{12}$CO $J=3$--2} \label{sec:co}

\begin{figure*}
    \centering
    \includegraphics[width=\linewidth]{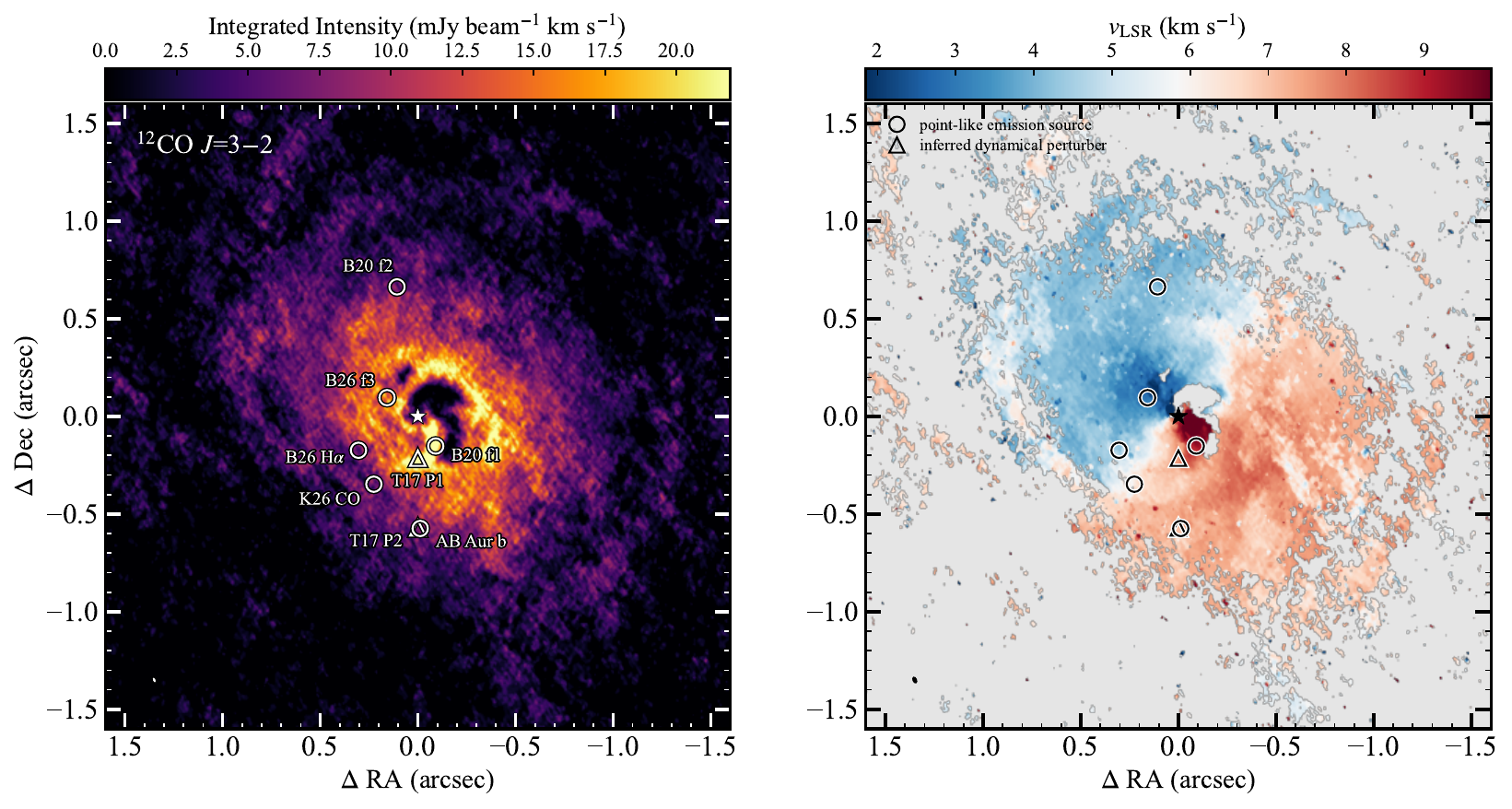}
    \caption{$^{12}$CO $J=3$--2 emission from the AB~Aur disk.
    \textit{Left:} integrated intensity (moment 0) map. \textit{Right:}
    intensity-weighted velocity (moment 1) map. The star symbol marks the stellar position, the ellipse south of the star marks the expected location of AB~Aur~b from \citet{Bowler2025}, and the synthesized beam is shown in the lower left corner of each panel. Open symbols mark the candidate planets within the field (Section~\ref{sec:ab_aur}): circles for point-like emission sources and triangles for inferred dynamical perturbers, following the classification of \citet{Speedie2024}. CO gas fills the continuum cavity and shows inner spiral substructure and a clear disk rotation pattern; the dark patches around the star are caused by spatial filtering of extended emission at velocities near $v_{\rm sys} \approx 5.85$\,km\,s$^{-1}$ \citep{tang2017}.}
    \label{fig:ab_aur_co}
\end{figure*}

Figure~\ref{fig:ab_aur_co} presents the $^{12}$CO $J=3$--2 integrated intensity (moment 0) and intensity-weighted velocity (moment 1) maps. To suppress noise, the moment maps are computed from the continuum-subtracted cube using a mask that retains pixels above $3\times$ the rms in each spatially smoothed channel.

In contrast to the continuum, the $^{12}$CO emission fills the central cavity: we detect gas from the outer disk down to the innermost regions, including spiral-like substructure interior to the dust ring (see also the channel maps in Appendix~\ref{sec:channel_map}), consistent with the inner CO spirals reported by \citet{tang2017} at lower angular resolution. The velocity field displays the clear pattern of a rotating disk, with the northern side blueshifted and the southern side redshifted about a systemic velocity of $v_{\rm sys} \approx 5.85$\,km\,s$^{-1}$ (LSR; \citealt{tang2017}).

At the location of AB~Aur~b we detect $^{12}$CO emission spanning $v_{\rm LSR} \approx 6$--8\,km\,s$^{-1}$, in line with the projected velocities of the surrounding disk gas at that position. We find no obvious localized compact excess or coherent velocity deviation associated with AB~Aur~b: the emission at its position is smoothly connected to the ambient disk in both intensity and velocity. The innermost candidates (B20~f1, B26~f3, and T17~P1) coincide with the bright inner CO spiral emission, whereas those at larger separations lie in lower-intensity regions of gas. None of the locations shows a brightness excess; all intensities lie within $2\sigma$ of the azimuthal median at the same radius.

\section{Discussion} \label{sec:discussion}

\begin{figure}
    \centering
    \includegraphics[width=\linewidth]{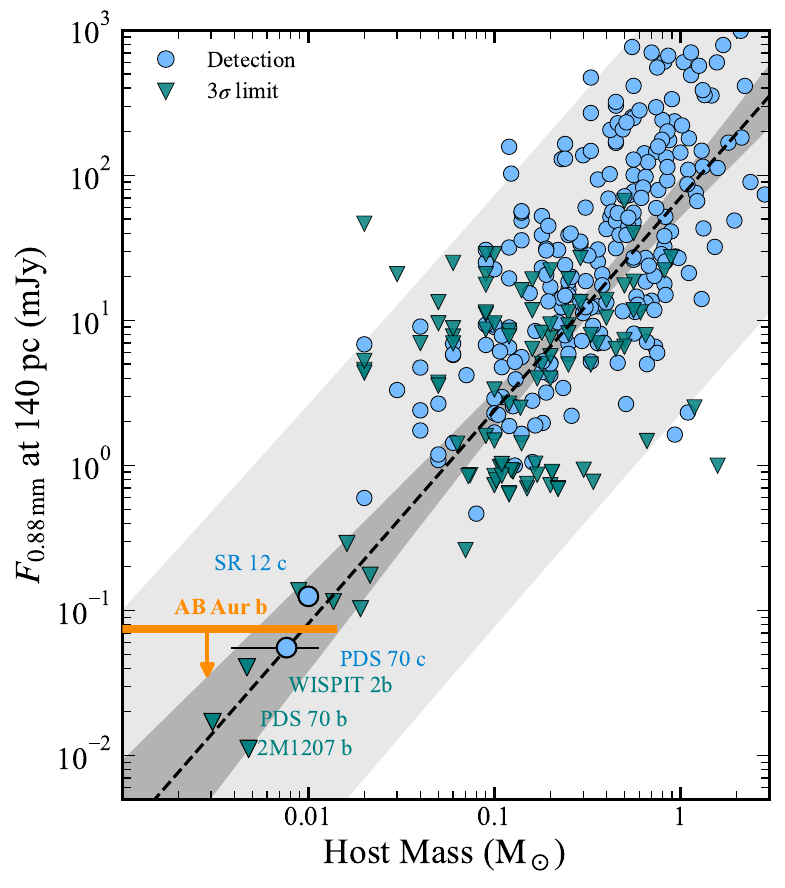}
    \caption{0.88\,mm flux density scaled to 140\,pc as a function of host mass (star, brown dwarf, or planetary-mass companion). Blue circles and teal triangles show detections and $3\sigma$ upper limits, respectively. Class~II disk measurements in Taurus, Lupus, and Chamaeleon~I and PMC limits are compiled from \citet{Wu2020}, with additional measurements for 2M1207~b \citep{Ricci2017}, WISPIT~2b \citep{Facchini2026}, PDS~70~b,c \citep{Benisty2021}, and SR~12~c \citep{Wu2022}. The black dashed line shows our Bayesian fit to the Class~II sample, including upper limits. Following \citet{Wu2020}, dark and light gray bands mark the 95\% credible interval for the mean relation and the 95\% predictive interval including 0.75~dex intrinsic scatter, respectively. The orange bar shows our $3\sigma$ limit for candidate CPDs in AB~Aur over an adopted companion mass range of 1--15\,$M_{\rm Jup}$.}
    \label{fig:dust_mass}
\end{figure}

Figure~\ref{fig:dust_mass} places our limit on sub-mm point sources around AB~Aur in the context of disks around stars, brown dwarfs, and planetary-mass companions (PMCs) in terms of the direct observable---the 0.88\,mm flux density scaled to a common distance of 140\,pc---which involves no assumptions on the dust opacity or temperature. The comparison sample comprises 0.88\,mm continuum measurements and $3\sigma$ upper limits for Class~II disks in Taurus, Lupus, and Chamaeleon~I, together with PMC disk limits for six wide, low-mass companions from \citet{Wu2020}. We also show the Band~7 detections of the disk around the planetary-mass object SR~12~c \citep{Wu2022} and the PDS~70~c CPD \citep{Benisty2021}, as well as upper limits for PDS~70~b \citep{Benisty2021}, WISPIT~2b \citep{Facchini2026}, and 2M1207~b \citep{Ricci2017}. For consistency, we include only measurements at 0.88\,mm as comparing observations at other wavelengths---such as the analysis of 1.3~mm sources from the DSHARP survey \citep{Andrews2021}---requires an assumption about the spectral index \citep[e.g.,][]{Testi2014, Ribas2017}.

To assess what an extrapolation from the stellar regime implies for the expected disk emission in the planetary regime, we fit the Taurus, Lupus, and Chamaeleon~I Class II samples compiled by \citet{Wu2020} using a Bayesian regression model \citep{Kelly2007}. We obtain \begin{equation} \log(F_{0.88\,{\rm mm}}/{\rm mJy}) = 1.85^{+0.13}_{-0.14} + \left(1.47^{+0.21}_{-0.20}\right) \log(M_\ast/M_\odot), \end{equation} with an intrinsic scatter of $0.75^{+0.08}_{-0.07}$~dex. The slope is consistent with the steeper-than-linear scaling found in previous studies \citep[e.g.,][]{Andrews2013, Pascucci2016}. Extrapolated to planetary masses, this relation predicts flux densities scaled to 140~pc of $0.027$, $0.076$, and $0.14$~mJy at 5, 10, and 15\,$M_{\rm Jup}$, respectively. Our sensitivity is comparable to the brightest known CPDs. In terms of dust mass, our limit ($0.0059$--$0.023\,M_\oplus$ for $T_{\rm dust}=40$--$15$~K; Section~\ref{sec:continuum}) is comparable to the inferred masses of the PDS~70~c CPD ($0.007$--$0.031\,M_\oplus$; \citealt{Benisty2021}) and the SR~12~c disk ($0.012$--$0.054\,M_\oplus$; \citealt{Wu2022}), with both ranges reflecting different assumed grain properties. Scaled to 140~pc, our $3\sigma$ limit of $0.074$~mJy is approximately 35\% higher than the PDS~70~c CPD flux of $0.055$~mJy \citep{Benisty2021}. Adopting the updated distance of $139\pm5$~pc to SR~12 \citep{finley2026}, an SR~12~c-like disk would have a scaled flux density of $0.125\pm0.016$~mJy and would be detected at ${\sim}5\sigma$. An OTS~44-like disk would be readily detectable at $0.45$~mJy (${\sim}18\sigma$), assuming a spectral index of 2.2 \citep{Wu2022}. Our observations therefore reach a depth comparable to the known CPDs: they exclude an SR~12~c-like disk at the $3\sigma$ level around AB~Aur~b and the other proposed candidates, and approach the flux of the PDS~70~c CPD. If the mass of AB~Aur~b is ${\lesssim}5\,M_{\rm Jup}$, our non-detection would be consistent with the deeper limits for WISPIT~2b, PDS~70~b, and 2M1207~b.

The physical interpretation of this non-detection depends on the optical depth of the emission. If the emission is optically thin, the flux limit can be converted into an upper limit on the dust mass. Optically thin emission is expected if the dust supply is low or grains are rapidly lost. Millimeter-sized grains can drift toward the planet on $10^2$--$10^3$\,yr timescales \citep{Zhu2018}, while size-selective inflow may limit their replenishment from the circumstellar disk \citep{Maeda2024}. A dust-evolution model tailored to AB~Aur~b predicts optically thin emission throughout the CPD at 1.3\,mm \citep{Shibaike2025}; the same model predicts Band~7 flux densities of ${\sim}10^2$\,$\mu$Jy for typical dust supply and weak turbulence, which our limit disfavors. In this regime, the inferred dust-mass limit also depends on the assumed opacity and temperature. A warmer CPD implies a lower limit, and a colder CPD a higher one.

CPDs may instead be optically thick. CPDs are expected to be truncated at roughly one third of the Hill radius \citep{Quillen1998}, a few au at most for AB~Aur~b and unresolved by our ${\sim}3$--5\,au beam. Accreting CPD models predict high dust surface densities and optically thick
emission over much of the disk \citep{Zhu2015, Zhu2019}. Dust traps can further build dense rings \citep{Drkazkowska2018}, and the reference model of \citet{Rab2019}, which assumes a standard dust-to-gas ratio, is optically thick at 860~$\mu$m over its inner ${\sim}3$\,au. In the optically thick limit, our flux limit constrains projected emitting area rather than dust mass: for a face-on, 40~K blackbody, the $3\sigma$ limit corresponds to an effective radius of ${\sim}0.41$\,au (or ${\sim}0.82$\,au at 15~K), well inside the expected truncation radius. This disfavors an extended optically thick emitting region under the blackbody assumption, but allows a more compact dust reservoir whose mass is essentially unconstrained.

The few existing CPD constraints illustrate this ambiguity. Compact, optically thick disks are among the proposed explanations for non-detections around wide-orbit planetary-mass companions \citep{Wu2017,Wu2020}, and remain a possible interpretation of the unresolved disk around SR~12~c \citep{Wu2022}. Both optically thin and thick cases were considered for PDS~70~c \citep{Benisty2021}. Subsequent observations found a millimeter spectral index of $\alpha=2.0\pm0.2$ across ALMA Bands~3--7, although a Band~9 non-detection falls below that extrapolation \citep{DominguezJamett2025}. The emission has since been modeled as an optically thick dust ring \citep{Shibaike2026} or as free--free emission \citep{DominguezJamett2025}. For future CPD detections, continuum observations at multiple wavelengths, together with a constraint on the emitting size, could help characterize optical depth and the emission mechanism.

Our non-detections do not, by themselves, disfavor a planetary interpretation for AB Aur b. A circumplanetary disk could remain undetected if it is sufficiently compact, optically thick, or depleted in millimeter-sized grains. At the same time, we find no independent compact continuum or localized CO signature associated with AB Aur b, so these observations provide no additional evidence for a circumplanetary structure at its position.

\section{Summary} \label{sec:summary}

We have presented the deepest (${\sim}20\,\mu$Jy\,beam$^{-1}$ rms) and highest-resolution (${\sim}30$~mas) ALMA Band~7 observations of AB~Aur to date, designed to search for CPDs around proposed protoplanets, including AB~Aur~b. Our main results are as follows:

\begin{enumerate}
\item We recover the 0.88\,mm dust ring peaking at ${\sim}0\farcs9$ (${\sim}140$\,au) with a strong azimuthal brightness asymmetry, and we detect compact emission at the stellar position which may be associated with free-free emission from a jet.
\item Apart from the central source, we detect no point source inside the cavity, down to a $3\sigma$ limit of $60\,\mu$Jy. Under standard optically thin assumptions it corresponds to a CPD dust mass of $\lesssim 0.0059\,M_\oplus$ ($T_{\rm dust}=40$\,K) or $\lesssim 0.023\,M_\oplus$ ($T_{\rm dust}=15$\,K), comparable to the PDS~70~c CPD.
\item $^{12}$CO $J=3$--2 emission fills the continuum cavity and shows inner spiral substructure and a clear disk rotation pattern. The CO emission at the position of AB~Aur~b is smoothly connected to the ambient disk, with no obvious localized excess or velocity deviation.
\item The millimeter non-detection implies that if AB~Aur~b or the other proposed candidate sources in this system are indeed protoplanets, their circumplanetary material is either lower in millimeter-emitting dust mass than the upper end of the PDS~70~c CPD range under comparable assumptions, very compact and optically thick, or depleted in large grains.
\end{enumerate}

\begin{acknowledgments}
B.P.B. acknowledges support from the National Science Foundation grant AST-1909209, NASA Exoplanet Research Program grant 20-XRP20$\_$2-0119, and the Alfred P. Sloan Foundation. We thank the North American ALMA Science Center (NAASC) for hosting a virtual
face-to-face data reduction visit in April 2025. We are especially grateful to
Arielle Moullet for scientific support throughout the visit, and to Holly Sheets
and Michael Sanchez for setting up the computing accounts and staging the data.
We also thank Catherine Vlahakis for assistance through the ALMA Helpdesk.

This paper makes use of the following ALMA data: ADS/JAO.ALMA\#2022.1.01360.S. ALMA is a partnership of ESO (representing its member states), NSF (USA) and NINS (Japan), together with NRC (Canada), NSTC and ASIAA (Taiwan), and KASI (Republic of Korea), in cooperation with the Republic of Chile. The Joint ALMA Observatory is operated by ESO, AUI/NRAO and NAOJ. The National Radio Astronomy Observatory is a facility of the U.S. National Science Foundation operated under cooperative agreement by Associated Universities, Inc.

\end{acknowledgments}
\clearpage
\bibliography{abaur_refs}{}
\bibliographystyle{aasjournalv7}



\appendix
\section{Channel Maps}\label{sec:channel_map}
\begin{figure*}[h]
    \centering
    \includegraphics[width=\linewidth]{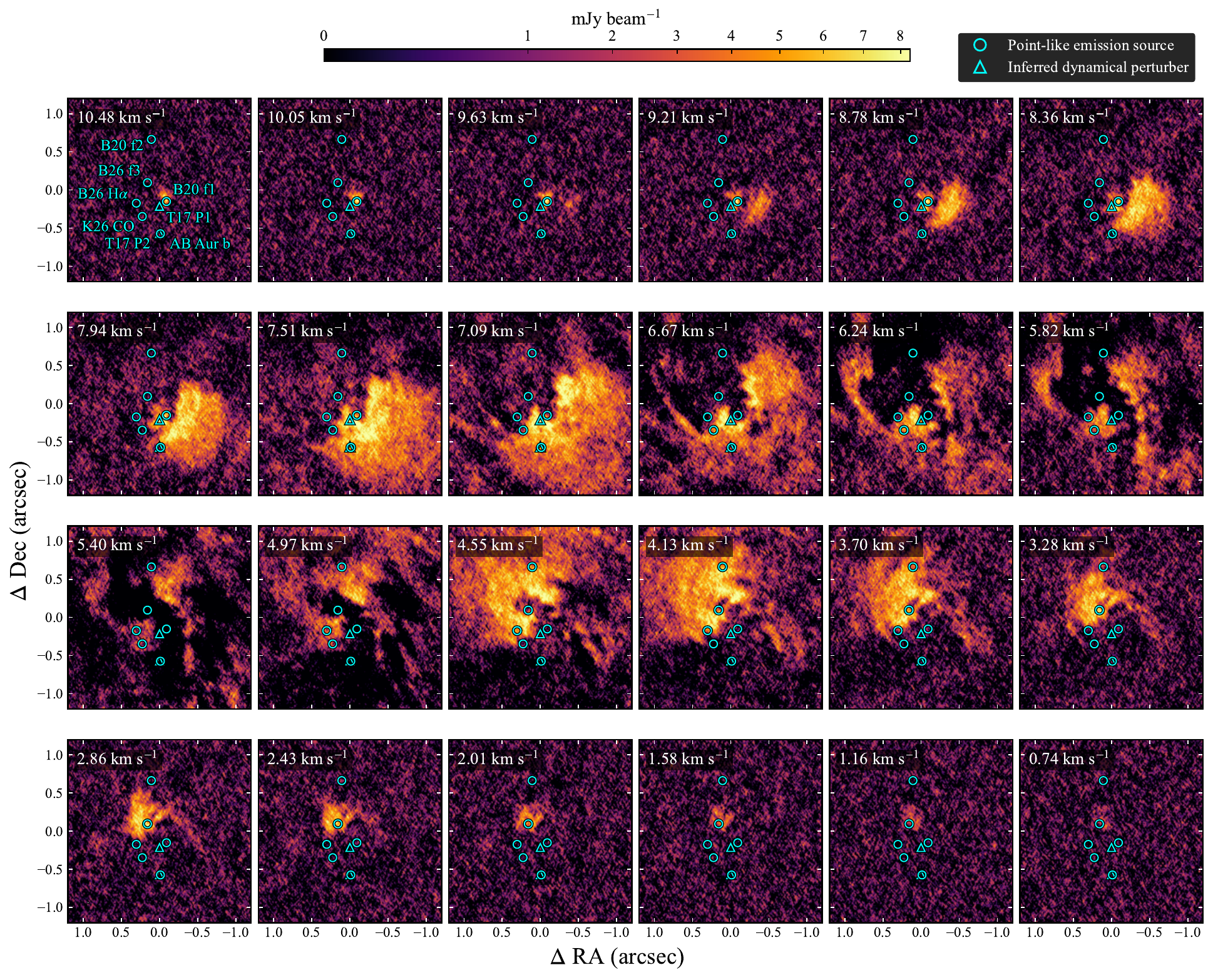}
    \caption{$^{12}$CO $J$=3--2 channel maps of AB~Aur at native resolution
    (beam $32\times19$~mas). The $v_{\rm LSR}$
    of each channel is shown in each panel. Cyan symbols mark the proposed
    planet-formation locations as in Figure~\ref{fig:ab_aur_co}; labels are
    shown in the first panel only.
    \label{fig:channels}}
    \label{fig:ab_aur_co_channel_map}
\end{figure*}

The $^{12}$CO channel maps (Figure~\ref{fig:channels}) show blueshifted
emission to the northeast and redshifted emission to the southwest, as
expected for disk rotation, with spiral arms on both sides within
${\sim}1\arcsec$ of the star, as seen in \citet{tang2017}. No velocity deviations are seen at the candidate locations at our channel width of 0.42~km~s$^{-1}$.

\end{document}